\documentclass[10pt,conference]{IEEEtran}
\IEEEoverridecommandlockouts
\usepackage{xurl} 
\usepackage[hidelinks]{hyperref}
\usepackage{breakurl}
\usepackage{cite}
\usepackage{amsmath,amssymb,amsfonts}
\usepackage{algorithmic}
\usepackage{subcaption}
\usepackage{graphicx}
\usepackage{float}
\pdfoutput=1
\usepackage[dvipsnames]{xcolor}
\usepackage{colortbl}
\usepackage[utf8]{inputenc}
\usepackage[english]{babel}
\usepackage[T1]{fontenc}

\usepackage{textcomp}

\usepackage{textcomp}
\usepackage{centernot}
\usepackage[braket, qm]{qcircuit}
\usepackage{braket}
\def\BibTeX{{\rm B\kern-.05em{\sc i\kern-.025em b}\kern-.08em
    T\kern-.1667em\lower.7ex\hbox{E}\kern-.125emX}}
\usepackage{multicol}
\usepackage{wrapfig}
\usepackage{tabularx}
\usepackage{tikz}
\usepackage{pgfplots} 
\usetikzlibrary{pgfplots.groupplots}
\usetikzlibrary{intersections}
\pgfplotsset{compat=1.16}
\pgfplotsset{scaled y ticks=false}
\usepgfplotslibrary{fillbetween}
\usepackage{mathtools}
\usepackage{listings}
\usepackage{booktabs} 

\usepackage{array}
\newcommand{\PreserveBackslash}[1]{\let\temp=\\#1\let\\=\temp}
\newcolumntype{C}[1]{>{\PreserveBackslash\centering}p{#1}}
\usepackage{comment}
\usepackage{multirow}
\usepackage{float}
\usepackage{cleveref}
\usepackage{algorithm2e}
\RestyleAlgo{ruled}

\definecolor{darkmagenta}{rgb}{0.55, 0.0, 0.55}
\usepackage{float}
\usepackage{afterpage}

\newif\ifdraft
\drafttrue
\ifdraft

\newcommand{\aanote}[1]{ {\textcolor{purple} { ***Abhishek: #1 }}}

\else
\newcommand{\note}[1]{}
\newcommand{\lpnote}[1]{}
\newcommand{\kwnote}[1]{}
\newcommand{\aanote}[1]{}
\newcommand{\mhnote}[1]{}
\fi

\begin{document}

\title{Real World Application of Quantum-Classical Optimization for Production Scheduling}
\author{
\IEEEauthorblockN{Abhishek Awasthi\IEEEauthorrefmark{1}, Nico Kraus\IEEEauthorrefmark{2}, Florian Krellner\IEEEauthorrefmark{3}, David Zambrano\IEEEauthorrefmark{2}}
\IEEEauthorblockA{\IEEEauthorrefmark{1}BASF Digital Solutions GmbH, Ludwigshafen am Rhein, Germany}
\IEEEauthorblockA{\IEEEauthorrefmark{2}Aqarios GmbH, Munich, Germany}
\IEEEauthorblockA{\IEEEauthorrefmark{3}SAP SE, Walldorf, Germany}
abhishek.awasthi@basf.com \\ 
\{nico.kraus, david.zambrano\}@aqarios.com \\ florian.krellner@sap.com}

\maketitle
 \thispagestyle{plain}
 \pagestyle{plain}

\begin{abstract}
This work is a benchmark study for quantum-classical computing method with a real-world optimization problem from industry. The problem involves scheduling and balancing jobs on different machines, with a non-linear objective function. We first present the motivation and the problem description, along with different modeling techniques for classical and quantum computing. The modeling for classical solvers has been done as a mixed-integer convex program, while for the quantum-classical solver we model the problem as a binary quadratic program, which is best suited to the D-Wave Leap’s Hybrid Solver. This ensures that all the solvers we use are fetched with dedicated and most suitable model(s). Henceforth, we carry out benchmarking and comparisons between classical and quantum-classical methods, on problem sizes ranging till approximately $\boldsymbol{150,000}$ variables. We utilize an industry grade classical solver and compare its results with D-Wave Leap’s Hybrid Solver. The results we obtain from D-Wave are highly competitive and sometimes offer speedups, compared to the classical solver.
\end{abstract}

\begin{IEEEkeywords}
Combinatorial Optimization, Quantum Annealing, D-Wave, Production \& Scheduling
\end{IEEEkeywords}

\section{Introduction}~\label{sec:introduction}

The Production Assignment and Scheduling problem is an NP-hard combinatorial optimization problem involving several jobs and machines, and focuses on balancing the processing time on the machines and reducing the manufacturing time of products while maximizing the value of the produced items. It arises in BASF agriculture business and is used to assign production jobs to chemical reactors optimally. Currently, the optimization process, i.e., finding an optimal or good enough solution, is a time-consuming process and any significant speed up would be highly welcome, for example, to allow more planning scenarios or larger planning horizons.

Combinatorial optimization problems, which involve finding the best solution among a finite set of possible options, are prevalent in various domains such as finance, logistics, and machine learning. Traditional optimization algorithms often struggle to efficiently solve combinatorial optimization problems due to the exponential growth in computational resources required as the problem size increases~\cite{Papadimitriou1981CombinatorialOA} and the general belief is that they cannot be solved efficiently. This limitation has led researchers to explore the potential of quantum computing in providing novel optimization techniques. 

The foundational principles of quantum mechanics, such as superposition and entanglement, enable quantum computers to perform calculations in parallel, potentially leading to exponential speed-up compared to classical computers for certain optimization problems~\cite{nielsen}. Quantum optimization algorithms, such as the quantum approximate optimization algorithm (QAOA) and the quantum annealing approach, have been developed to harness this quantum advantage~\cite{farhi,Albash_2018}. One of the key applications of quantum optimization is in the field of combinatorial optimization. Several experimental platforms have been developed to implement quantum optimization algorithms. Quantum annealers, such as those based on superconducting qubits or neutral atoms, utilize the principles of quantum annealing to find the optimal solution by minimizing the energy of a quantum system~\cite{Johnson2011Quantum}. On the other hand, gate-based quantum computers, which use quantum logic gates to manipulate qubits, offer greater flexibility but face challenges in terms of noise and error rates~\cite{Preskill_2018}. 

However, translating this potential into a practically relevant quantum advantage has proven to be a very challenging endeavor and no practical quantum advantage or advantage by using quantum computing has been found for combinatorial optimization problems relevant in industry, for example, recently~\cite{Leib2023, Awasthi_2023} could not show any advantage by using quantum computing. At the current stage of quantum hardware, no advantage can be achieved by using pure quantum methods to solve combinatorial optimization problems, due the quantum hardware is only able to solve small optimization problems not relevant in industry. Hence, it is only possible to see any practical advantages by using quantum computing when using quantum-classical methods, which leverage both classical and quantum computing.

Our problem can also be interpreted as vehicle routing problem (VRP)~\cite{vrp_book} with the following translation: Machines to vehicles, jobs to cities, and setup times to distance between cities. Earlier studies on solving the VRP with quantum computing are~\cite{feld2019VRP_quantum, iccs2020VRP_quantum} and more recently~\cite{sinno2023performancecommercialquantumannealing}.

In this work we compare the performance of the best classical solution method Gurobi~\cite{gurobi} with the D-Wave Leap’s Hybrid Solver~\cite{dwave_solver}, with the results being presented in Section~\ref{sec:results}. Since the performance of the solvers depending on how the model is formulated, we introduce two different but equivalent formulations in the next section. For the classical solver, we formulate the problem as a mixed-integer convex program with a small number of quadratic terms in the objective. On the other hand, the quantum-classical solver can deal very well with quadratic terms but does not support non-binary variables (natively). Hence, for the quantum-classical method we formulate the problem as a pure binary quadratic optimization problem.

Our analyses show that the D-Wave Leap’s Hybrid Solver outperforms the classical solver Gurobi on our test set, whereas it is not known if it is due to utilizing quantum computing. Nevertheless, it either shows that it is possible to develop classical general-purpose heuristics for quadratic binary optimization problems that can enhance classical state-of-the-art exact mixed-integer programming solvers such as Gurobi, or combinatorial optimization can already benefit from quantum annealing.

\section{Problem description and mathematical models}~\label{sec:modeling}

In this section we introduce the problem and present dedicated mathematical models most suited to classical and quantum computing solvers, each. We first present a mixed-integer convex program that has a small number of quadratic terms in the objective. Then we model the problem as a pure binary quadratic program suited for the quantum-classical solution method we use.

\subsection{Problem definition}
\begin{figure*}[ht]
 \centering
 \includegraphics[width=\linewidth]{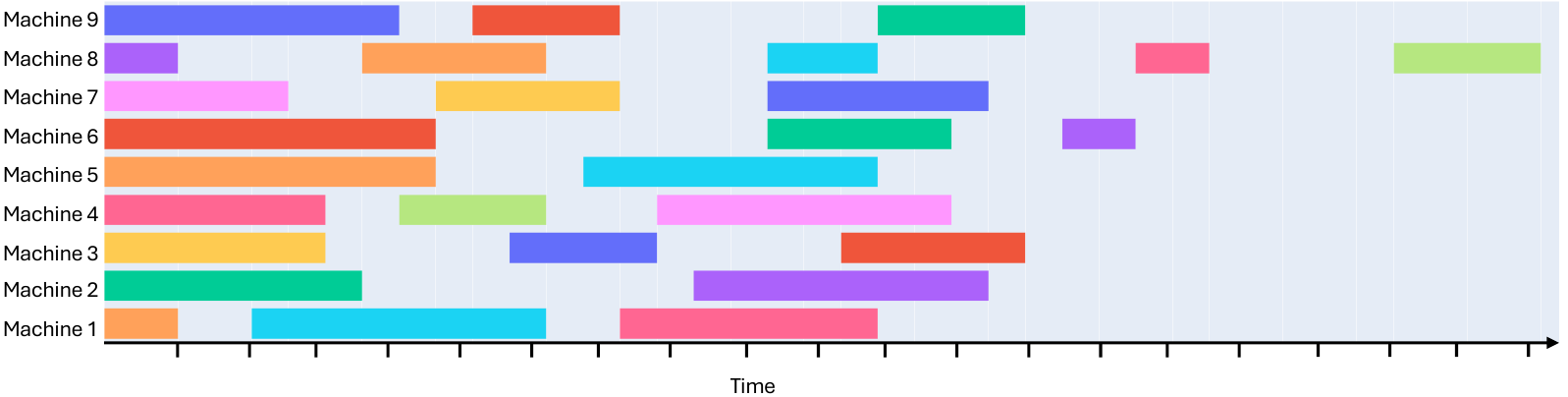}
 \caption{An exemplary solution to a problem with $27$ jobs and $9$-machines. All the machines can start processing from time $t=0$ and need to respect a setup time for each subsequent job. The objective is to reduce the total setup time, uniformly distribute the busy times of all machines, and maximize the total value.}
 \label{fig0}
\end{figure*}

The Production Assignment and Scheduling problem is a mathematical optimization problem involving several jobs and machines, and focuses on balancing the processing time on the machines and reducing the manufacturing time of products while maximizing the value of the produced items. Formally, we are given a set of machines $\mathcal{M} \in \{1,2,\cdots,M\}$ which process different jobs $\mathcal{J}\in \{ 1,2,\dots, J\}$. Each job $j\in \mathcal{J}$ requires a machine specific processing time $p_j^m$ for the machine $m \in \mathcal{M}$. Additionally, not all the jobs can be processed by all the machines. Hence, we are provided with a set of eligible machines $\mathcal{E}_j \subseteq \mathcal{M}$ for each job $j \in \mathcal{J}$. Finally, it is also required that each machine needs to be prepared (cleaned or setup) to process a job. This setup is accounted for in terms of time, and the setup time is dependent on the immediate previous job processed by the machine. These setup times are machine independent. Suppose, machine $m$ processes job $i \in \mathcal{J}$ and immediately followed by job $j \in \mathcal{J}$, then the setup time required for machine $m$ to start the processing of job $j$ is denoted as $s_{i,j}$. Successful processing of a job $j$ on a machine $m$ brings in a value of $v_{j,m}$, which is desired to be maximized. Figure~\ref{fig0}, shows an exemplary solution as a Gantt chart for a problem with $27$ jobs and $9$ machines.

\subsection{Mixed-integer convex programming formulation}~\label{sec:relaxation}

The formulation presented in this section is a modification of a mathematical programming formulation of the well-known vehicle routing problem and for one machine it is a mixed-integer linear programming formulation of the travel salesperson problem.

The variables defining the sequence of jobs on each machine are
\begin{equation}
x_{m,i,j}=
\begin{cases}
 1, & \text{if job } j \text{ is right after job } i \text{ on machine } m \\
 0, & \text{otherwise}
\end{cases}
\end{equation}
for all machines $m\in \mathcal{M}$ and jobs $i,j \in \mathcal{J}_0 := \mathcal{J} \cup \{0\}$ with $i \neq j$. Job $0$ is a dummy job that is the predecessor of every first job $j_1 \in \mathcal{J}$ scheduled on a machine and the successor of the last job $j_2 \in \mathcal{J}$ on the machine. All parameters corresponding to the dummy job are zero.

To model the problem we need two more types of variables. The variables $u_{m,j} \geq 0$ represent the cumulative processing time of machine $m \in \mathcal{M}$ right after job $j\in \mathcal{J}$, if $j$ is placed on machine $m$. These variables are used to disallow cycles in the predecessor relations. For example, it is not allowed to have job $1$ be the direct predecessor of job $2$, job $2$ is the direct predecessor of job $3$ and this job is the direct predecessor of job $1$. The variables $t_m \geq 0$ are the total processing time of the jobs placed on machine $m \in \mathcal{M}$.

The jobs are placed on machines by assigning a direct predecessor and successor to them. Each job $j \in \mathcal{J}$ needs exactly one predecessor, that is,
\begin{equation}
\sum_{m \in \mathcal{M}}\sum_{i \in \mathcal{J}_0} x_{m,i,j} = 1
\end{equation}
and each job $i \in \mathcal{J}$ has exactly one successor:
\begin{equation}
\sum_{m \in \mathcal{M}}\sum_{j \in \mathcal{J}_0} x_{m,i,j} = 1.
\end{equation}

To ensure that the direct predecessor and the successors are on the same machine we require:
\begin{equation}
\sum_{i \in \mathcal{J}_0} x_{m,i,j} - \sum_{k \in \mathcal{J}_0} x_{m,j,k} = 0
\end{equation}
for each machine $m \in \mathcal{M}$ and job $j \in \mathcal{J}_0$.

The dummy job $0$ can at most have one successor job $j \in \mathcal{J}_0$ for each machine $m \in \mathcal{M}$:
\begin{equation}
\sum_{j \in \mathcal{J}} x_{m,0,j} \leq 1.
\end{equation}

To disallow cycles in the direct predecessor and successor relations we need the following constraints
\begin{equation}
u_{m,i} - u_{m,j} + \mathtt{M} \cdot x_{m,i,j} \leq \mathtt{M} -p_j^m
\end{equation}
for each machine $m \in \mathcal{M}$ and job $j\in \mathcal{J}$. $\mathtt{M} $ is a large positive number, for example, the sum of all processing times. The constraints are known as the generalized sub-tour elimination constraints in the vehicle routing domain. Notice, that if $x_{m,i,j}=1$, meaning that job $j$ is the direct predecessor of job $i$ on machine $m$, the equation reads as
\begin{equation}
u_{m,i} + p_j^m \leq u_{m,j},
\end{equation}
stating that $u_{m,j}$ must be at least the cumulative processing time of all the jobs before and including job $j$. 

To ensure that $t_m$ is actually the cumulative processing time of machine $m \in \mathcal{M}$, we require
\begin{equation}
t_m \geq u_{m,j}
\end{equation}
for all jobs $j \in \mathcal{J}$.

The following cutting plane ensures a tight relaxation of the problem:
\begin{equation}
\sum_{m \in \mathcal{M}} t_m = \sum_{j \in \mathcal{J}} p_j^m.
\end{equation}
In words, the sum of the cumulative processing times for each machine is the sum of all processing times.

If there are combinations of jobs $j \in \mathcal{J}$ and machines with $m \in \mathcal{M}$ that are not allowed, we set
\begin{equation}
x_{m,i,j} = 0.
\end{equation}
for all $i \in \mathcal{J}_0$.

The objective function is the composition of three different optimization objectives. First, we want to minimize the total setup time on all the machines.
\begin{equation}
\min \sum_{m \in \mathcal{M}} \sum_{i \in \mathcal{J}} \sum_{j \in \mathcal{J}} s_{i,j}\cdot x_{m,i,j}.
\end{equation}

Second, we prefer uniform busy times for all the machines. That means that the sum of the processing times on the machines should be as equal as possible, i.e.,
\begin{equation}
\min \sum_{m \in \mathcal{M}} t_m^2.
\end{equation}

Last, we want to maximize the overall value of the processed jobs:
\begin{equation}
\max \sum_{m \in \mathcal{M}} \sum_{i \in \mathcal{J}_0} \sum_{j \in \mathcal{J}} v_{j,m} \cdot x_{m,i,j}.
\end{equation}

Notice, that the relaxation of the problem, i.e., relax $x_{m,i,j} \in \{0,1\}$ to $x_{m,i,j} \in [0,1]$, is a convex problem. In addition, the formulation has a minimal number of quadratic objective terms; one for each machine.


\subsection{Binary quadratic optimization model}~\label{sec:quadratic}

The drawback of the mixed-integer convex programming formulation for quantum computing is that it needs continuous (or integer) variables. Current quantum optimization method naturally support only binary variables. Thus the continuous (or integer) variables must ideally be converted to binary variables. In this section we present a pure binary formulation that is equivalent to the formulation in Section~\ref{sec:relaxation}. Specifically, there is a bijection between the variables with identical objective values. The model presented in this section has roughly the same number of binary variables as the MIP model.

In this formulation, we assign each job to a position in an order corresponding to the machines, similar to the quadratic programming formulation of the vehicle routing problem. We define the binary decision variables as,

\begin{equation}
x_{m,t,j} = \begin{cases}
1, &\text{if job $j$ is the $t$-th job on machine $m$}\\
0, &\text{otherwise}
\end{cases}
\end{equation}
for all machines $m \in \mathcal{M}$, times $t \in \mathcal{T}$, and jobs $j \in \mathcal{J}$. Here, $\mathcal{T}=\{1,\ldots,J\}$ are the possible positions of a job in a job-sequence on a machine.

The constraints of the model ensure that every job is processed at least once, that the $x_{m,t,j}$ actually define a sequence for each machine $m$ and that at most one job can be processed by any machine at any given time. 

That all jobs are scheduled on some machine, once and only once, is ensured by
\begin{equation}
\sum_{m \in \mathcal{M}}\sum_{t \in \mathcal{T}} x_{m,t,j} = 1
\end{equation}
for each job $j\in \mathcal{J}$. 

That at most one job can be processed by any machine at any given time, is ensured by
\begin{equation}
\sum_{j \in \mathcal{J}} x_{m,t,j} \leq 1
\end{equation}
for all machines $ m\in \mathcal{M}$ and $t\in \mathcal{T}$. 

To ensure a valid order of jobs on each machine, a job can only be scheduled at $t\in \{2,3,\dots,|\mathcal{T}|\}$ on machine $m \in \mathcal{M}$, if another job is scheduled at time step $t-1$. As a formula:
\begin{equation}
\sum_{j \in \mathcal{J}} x_{m,t,j} \leq \sum_{j \in \mathcal{J}} x_{m,t-1,j}.
\end{equation}
Formulated differently, the constraints prevent the following assignment: $x_{m,t-1,j_1}=0$ and $x_{m,t,j_2}=1$ for all jobs $j_1 \in \mathcal{J}$, some job $j_2 \in \mathcal{J}$, and $t \geq 2$. 

As before, if job $j\in \mathcal{J}$ cannot be processed on a machine $m \notin \mathcal{E}_j$, then $x_{m,t,j} = 0$ for all $t\in \mathcal{T}$.


The objective is as in Section~\ref{sec:relaxation}. We minimize the total setup times on all the machines
\begin{equation}
\min \sum_{m \in \mathcal{M}} \sum_{t \in \mathcal{T}^\prime} \sum_{i \in \mathcal{J}} \sum_{j \in \mathcal{J}} s_{i,j} \cdot x_{m,t-1,i} \cdot x_{m,t,j},
\end{equation}
utilize the machines as equal as possible
\begin{equation}
\min \sum_{m \in \mathcal{M}} \left( \sum_{t \in \mathcal{T}} \sum_{j \in \mathcal{J}} p_j^m \cdot x_{m,t,j}\right)^2,
\end{equation}
and maximize the overall value of processing the jobs
\begin{equation}
\max \sum_{m \in \mathcal{M}} \sum_{t \in \mathcal{T}} \sum_{j \in \mathcal{J}} v_{j,m} \cdot x_{m,t,j}
\end{equation}
by linearly combining the three objectives.





\section{Experiments and results}~\label{sec:results}

\begin{figure*}
\centering
\begin{minipage}{1\columnwidth}
\centering
\includegraphics[width=\textwidth]{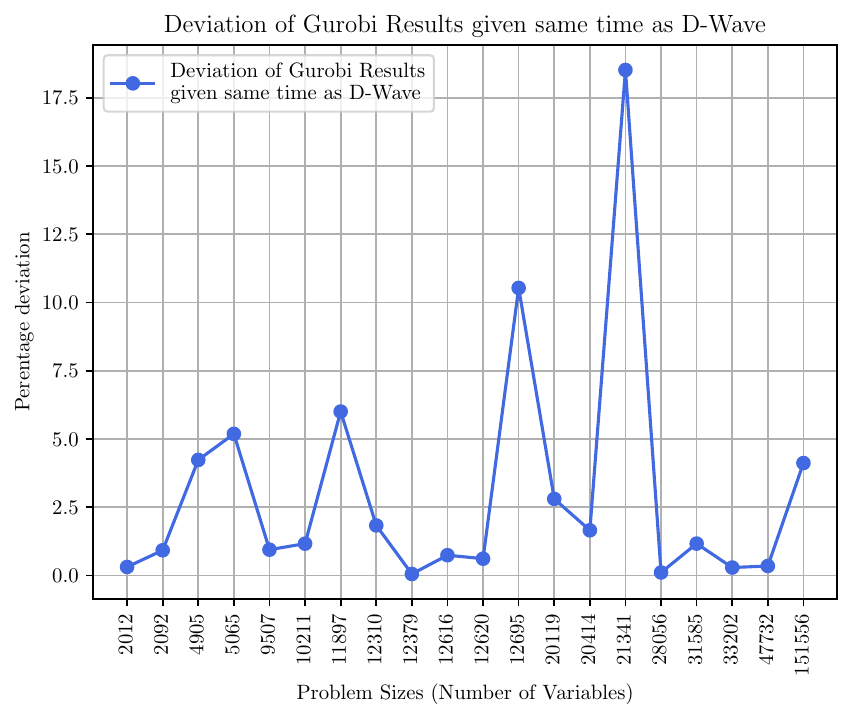}
\caption{\small{Percentage deviation of Gurobi results for all the instances, given the same runtime as D-Wave.}}
\label{fig3}
\end{minipage}%
\quad
\begin{minipage}{1\columnwidth}
\centering
\includegraphics[width=\textwidth]{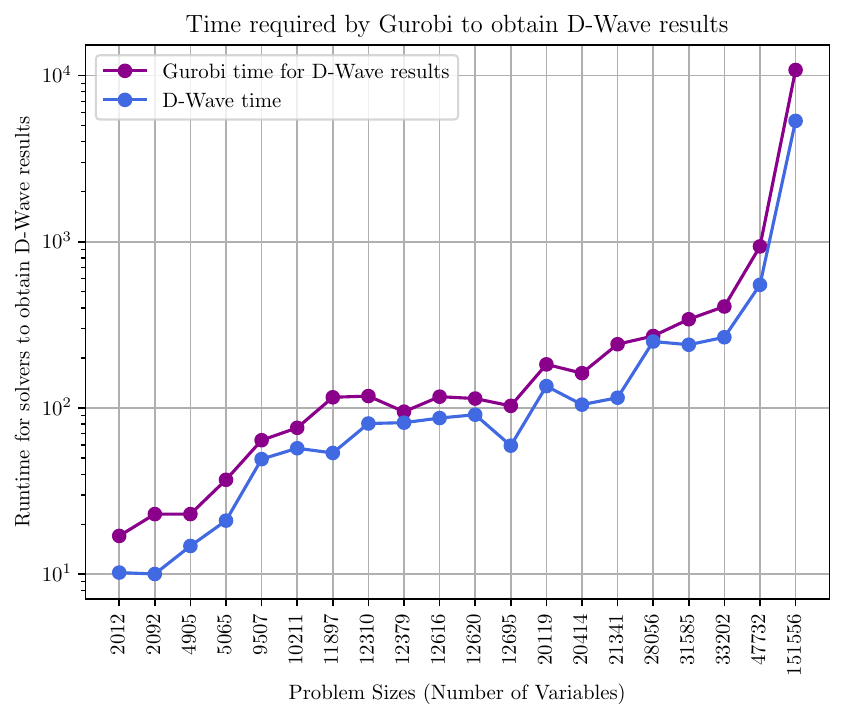}
\caption{\small{Runtime required by Gurobi and D-Wave, to obtain the same solution quality as D-Wave.}}
\label{fig4}
\end{minipage}
\end{figure*}

\subsection{Optimization solvers}

Gurobi is a (powerful) mathematical programming solver that utilizes the branch and bound algorithm. It is considered state-of-the-art and is designed to run on classical hardware. This algorithm works by breaking down the original problem into smaller sub-problems using a decision tree. By using lower bounds derived from the continuous relaxations, Gurobi is able to reduce the search space and improve efficiency~\cite{gurobi}. To do this efficiently, Gurobi must be able to solve the relaxation efficiently, which is the case for a convex relaxation (with a small number of quadratic terms in the objective). Gurobi is an exact and all-purpose solver, meaning it can be used to solve a wide range of optimization problems. Exact means that Gurobi is capable of finding a guaranteed optimal solution in a deterministic manner, given enough time. 

D-Wave provides a pure quantum as well as a quantum-classical solver to solve optimization problems. D-Waves quantum annealer has over $5,000$ qubits and an average of $15$ couplers per qubit \cite{farre}. The quantum-classical solver D-Wave Leap’s Hybrid solver (from now on D-Wave) combines the quantum annealer technology with classical algorithms. The quantum-classical solver allows solving larger optimization problems that exceed the capacity of the quantum hardware by decomposing the original problem into smaller parts and presenting only a subset of the problem to the quantum annealer. However, the exact workings of the method are not publicly disclosed.

\begin{figure*}[t]
\centering
\begin{minipage}{1\columnwidth}
\centering
\includegraphics[width=\textwidth]{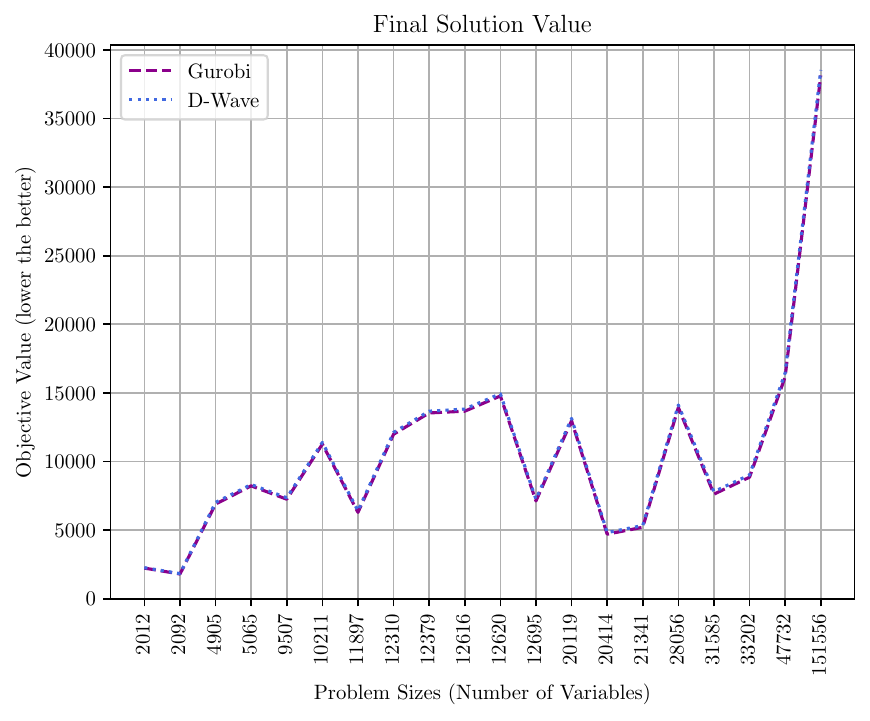}
\caption{\small{Final objective value obtained from the two solvers, for different problem sizes. The plots show that the solution quality is almost identical.}}
\label{fig1}
\end{minipage}%
\quad
\begin{minipage}{1\columnwidth}
\centering
\includegraphics[width=0.98\textwidth]{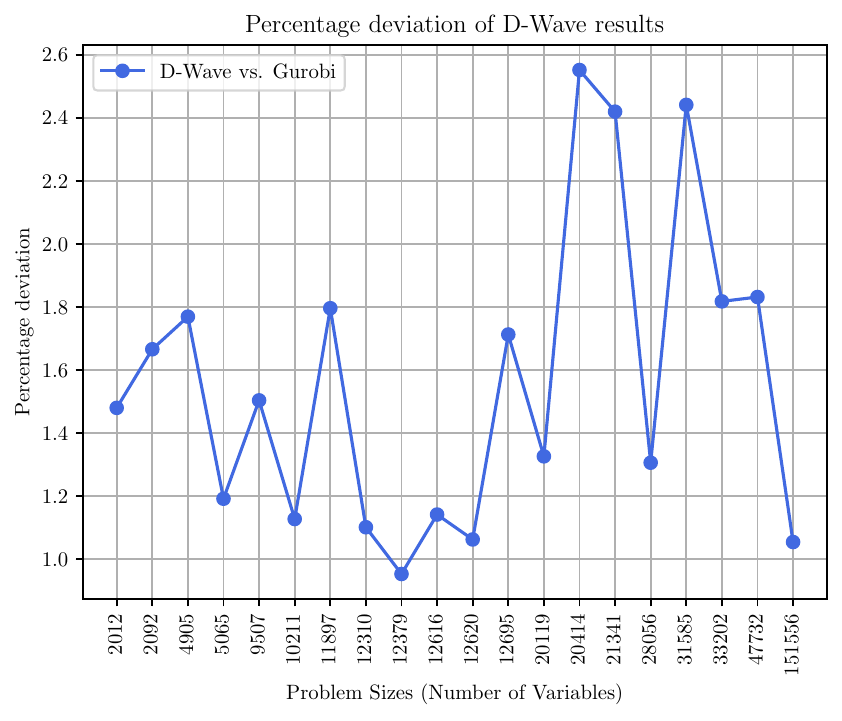}
\caption{\small{Percentage deviation of the D-Wave final objective value from the final solution of Gurobi. We can see that the worst solution provided by D-Wave is within $2.6\%$ to that of Gurobi solver.}}
\label{fig2}
\end{minipage}
\end{figure*}

\subsection{Experimental setup}

\begin{table}[!htbp]
\centering
\begin{tabular}{c|c|c|c|c} \hline\hline
 \multirow{2}{*}{\textbf{Jobs}} & \multirow{2}{*}{\textbf{Machines}} & \textbf{D-Wave} & \textbf{Gurobi} & \textbf{D-Wave} \\ 
 & & \textbf{Variables} & \textbf{Variables} & \textbf{Runtime (seconds)} \\ \hline 
27 & 9 & 2,012 & 2,432 & 5.00 \\ \hline
27 & 9 & 2,092 & 2,518 & 5.00 \\ \hline
45 & 9 & 4,905 & 5,550 & 7.39 \\ \hline
48 & 8 & 5,065 & 5,684 & 8.48 \\ \hline
50 & 10 & 9,507 & 10,448 & 23.52 \\ \hline
60 & 10 & 10,211 & 11,182 & 27.71 \\ \hline
60 & 15 & 11,897 & 13,184 & 25.68 \\ \hline
63 & 9 & 12,310 & 13,324 & 42.32 \\ \hline
63 & 9 & 12,379 & 13,396 & 42.75 \\ \hline
63 & 9 & 12,616 & 13,642 & 44.40 \\ \hline
72 & 9 & 12,620 & 13,640 & 46.59 \\ \hline
60 & 15 & 12,695 & 14,024 & 29.06 \\ \hline
75 & 15 & 20,119 & 21,790 & 68.89 \\ \hline
60 & 20 & 20,414 & 22,362 & 54.24 \\ \hline
60 & 20 & 21,341 & 23,334 & 58.70 \\ \hline
90 & 15 & 28,056 & 30,024 & 131.66 \\ \hline
80 & 20 & 31,585 & 33,998 & 127.22 \\ \hline
80 & 20 & 33,202 & 35,672 & 140.50 \\ \hline
100 & 20 & 47,732 & 50,694 & 281.87 \\ \hline
160 & 20 & 151,556 & 156,810 & 2744.98 \\ \hline \hline
\end{tabular}
\caption{Problem instances used for benchmarking in this work. Gurobi was provided with a constant runtime of three hours.}
\label{table_1}
\end{table}

The problem sizes we study are ranging from $2,012$ to $151,556$ variables for the binary quadratic formulation and slightly more variables are needed for the quadratic mixed-integer convex program, as shown in Table~\ref{table_1}. 

D-Wave was provided with a specific solve-time depending on the problem size and the number of constraints in the problem~\cite{dwave_timer}, as shown in Table~\ref{table_1}. Gurobi was provided with a time limit of $3$-hours, and the hardware utilized to run the Gurobi solver was an Intel$^\text{{\textregistered}}$ Xeon$^\text{{\textregistered}}$ Platinum 8124M CPU with $3$GHz, consisting of 8 physical cores and 16 logical processors.


To compare the performance of the two solvers, we compare the solution quality reached by Gurobi given the same runtime as D-Wave for each instance as shown in Figure~\ref{fig3} and we compare the runtime required by Gurobi to obtain the same solution quality as D-Wave, shown in Figure~\ref{fig4}. We can see that given the same runtime as D-Wave, the solution quality of Gurobi was worse for several instances, with a maximum deviation of more than $17.5\%$ to the D-Wave results. The second analysis suggests that the D-Wave solver offers a speed-up of approximately $1.5$ to $2$, compared to Gurobi for the investigated instances. 

We also compare the solution quality obtained from both the solvers, provided the runtimes explained above. Figure~\ref{fig1} shows the final objective value obtained from both the solvers. It can be observed that the solution value from both the solvers converges to same values, for all the instances tested in this work. Figure~\ref{fig2} shows the exact percentage deviation of D-Wave results to the final solution value from Gurobi. The performance of D-Wave is extremely good, given the fact that Gurobi was provided with a runtime of $3$ hours, while, D-Wave was provided with a maximum of $45$ minutes for the largest problem with $151,556$ variables.

\begin{figure}
\centering
\includegraphics[width=0.45\textwidth]{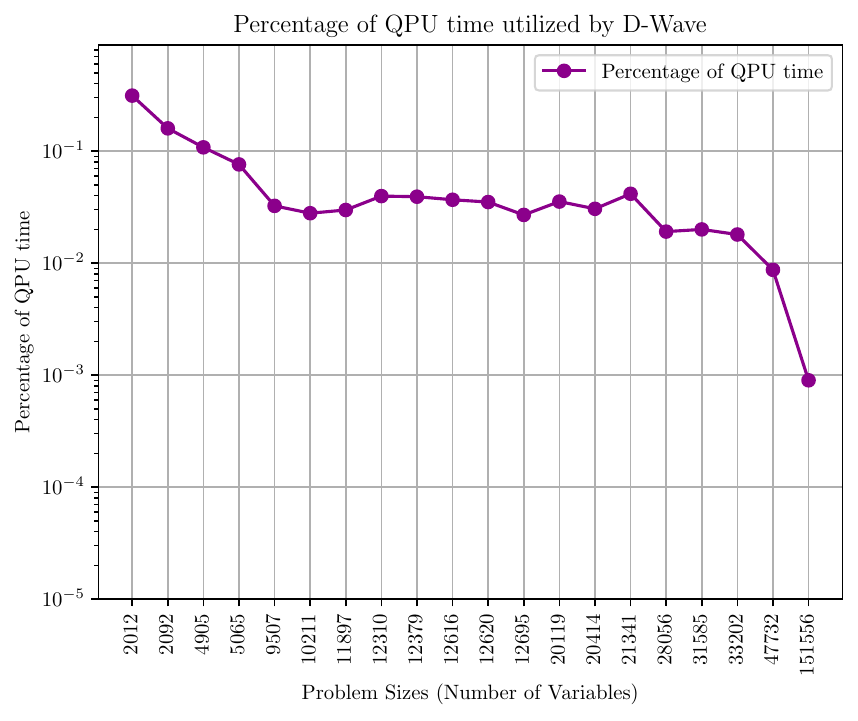}
\caption{\small{Percentage of QPU time utilized by D-Wave solver.}}
\label{fig5}
\end{figure}

\section{Conclusion}~\label{sec:conclusion}
In this work we attempt to compare a classical and quantum-classical solver for an industrially relevant optimization problem. We attempt a fair comparison of the two solvers, firstly, by modeling the problems in the most suitable manner for each solver, respectively, and secondly, by analyzing the solution quality and the runtimes given a fixed amount of runtime and solution quality. In our experiments, it can be seen that the results provided by D-Wave are of competitive quality and even offer speedups for larger problem instances. However, we would also like to highlight that D-Wave is a black-box quantum-classical solver, and it is not entirely clear how the classical solvers/resources are combined with the quantum-chip, to solve such large problem instances. Figure~\ref{fig5} shows the percentage of QPU time utilized by the D-Wave solver for all the problems in this work. Nonetheless, the solution quality and the speedup offered by D-Wave solver are encouraging signs that there exist optimization problems where can expect some speedup using quantum-classical solvers.

\section*{Acknowledgement}
We would like to acknowledge the Federal Ministry for Economic Affairs and Climate Action (abbreviated BMWK), for funding this work under the QCHALLenge Project (01MQ22008A). We thank Sarah Braun for discussions and proof-reading the manuscript. Finally, we would also like to acknowledge Alexander Milczarek and Ron Lehmann from BASF SE, for providing us with the studied optimization use-case. 



\end{document}